\documentclass{article}

\PassOptionsToPackage{numbers, compress}{natbib}
\usepackage[preprint]{neurips_2023}

\usepackage[utf8]{inputenc} % allow utf-8 input
\usepackage[T1]{fontenc}    % use 8-bit T1 fonts
\usepackage{hyperref}       % hyperlinks
\usepackage{url}            % simple URL typesetting
\usepackage{booktabs}       % professional-quality tables
\usepackage{amsfonts}       % blackboard math symbols
\usepackage{nicefrac}       % compact symbols for 1/2, etc.
\usepackage{microtype}      % microtypography
\usepackage{xcolor}         % colors

\usepackage{amsmath,amssymb,amsfonts}
\usepackage{algorithmic}
\usepackage{graphicx}
\usepackage{textcomp}
\usepackage{graphicx}
\usepackage{caption}
\usepackage{subcaption}
\usepackage{bbm}
\usepackage{bm}
\usepackage{color}
\usepackage{hyperref}
\usepackage{multirow}
\usepackage{enumitem}
\usepackage{algorithm} 
\usepackage{algorithmic}
\usepackage{enumerate}
\usepackage{amsmath}
\usepackage{booktabs}
\usepackage{multirow}
\usepackage{stmaryrd}
\usepackage{color}
\usepackage{balance}
\usepackage{adjustbox}
\usepackage{mathtools}

\newcommand{\etal}{\textit{et al}. }
\newcommand{\etc}{\textit{etc}.}
\newcommand{\eg}{\textit{e.g.}}

\newcommand{\ie}{\emph{i.e.}}

\title{Play to Your Strengths: Collaborative Intelligence of Conventional Recommender Models and Large Language Models}

\author{%
    Yunjia Xi \\
    Shanghai Jiao Tong University \\
    Shanghai, China \\
    xiyunjia@sjtu.edu.cn \\
    \And 
    Weiwen Liu  \\
    Huawei Noah’s Ark Lab \\
    Shenzhen, China \\
    liuweiwen8@huawei.com \\
    \And 
    Jianghao Lin  \\
    Shanghai Jiao Tong University \\
    Shanghai, China \\
    chiangel@sjtu.edu.cn \\
    \And 
    Chuhan Wu  \\
    Huawei Noah’s Ark Lab \\
    Beijing, China \\
    wuchuhan1@huawei.com \\
    \And 
    Bo Chen \\
    Huawei Noah’s Ark Lab \\
    Shenzhen, China \\
    chenbo116@huawei.com \\
    \And 
    Ruiming Tang \\
    Huawei Noah’s Ark Lab \\
    Shenzhen, China \\
    tangruiming@huawei.com \\
    \And 
    Weinan Zhang \\
    Shanghai Jiao Tong University \\
    Shenzhen, China \\
    wnzhang@sjtu.edu.cn \\
    \And 
    Yong Yu \thanks{Corresponding author.} \\
    Shanghai Jiao Tong University \\
    Shanghai, China \\
    yyu@sjtu.edu.cn \\
}

\begin{document}

\maketitle

\begin{abstract}
The rise of large language models (LLMs) has opened new opportunities in Recommender Systems (RSs) by enhancing user behavior modeling and content understanding. However, current approaches that integrate LLMs into RSs solely utilize either LLM or conventional recommender model (CRM) to generate final recommendations,
without considering which data segments LLM or CRM excel in. To fill in this gap, we conduct experiments on MovieLens-1M and Amazon-Books datasets, and compare the performance of a representative CRM (DCNv2) and an LLM (LLaMA2-7B) on various groups of data samples. Our findings reveal that LLMs excel in data segments where CRMs exhibit lower confidence and precision, while samples where CRM excels are relatively challenging for LLM, requiring substantial training data and a long training time for comparable performance. This suggests potential synergies in the combination between LLM and CRM. Motivated by these insights, we propose \underline{Co}llaborative Recommendation with conventional \underline{Re}commender and \underline{L}arge \underline{La}nguage Model (dubbed \textit{CoReLLa}). In this framework, we first jointly train LLM and CRM and address the issue of decision boundary shifts through alignment loss. Then, the resource-efficient CRM, with a shorter inference time, handles simple and
moderate samples, while LLM processes the small subset of challenging samples for CRM. Our experimental results demonstrate that CoReLLa outperforms state-of-the-art CRM and LLM methods significantly, underscoring its effectiveness in recommendation tasks.
\end{abstract}

\section{Introduction}
% Recently, the rise of large language models (LLMs) has opened up new opportunities in the field of Recommender Systems (RSs). With their plentiful world knowledge and reasoning ability, LLMs can enhance user behavior modeling and content understanding in RSs, leading to more precise and tailored recommendations~\cite{lin2023can,xi2023towards}. However, current works that integrate LLMs into RSs solely utilize either LLM or conventional recommender model (CRM) to generate final recommendations, without considering which data segments LLM or CRM really excel in. 

In recent years, the emergence of large language models (LLMs) has opened up new opportunities within the realm of Recommender Systems (RSs). These LLMs, with their vast array of world knowledge and sophisticated reasoning capabilities, offer a unique opportunity to revolutionize user behavior modeling and content comprehension within RSs, thereby facilitating the delivery of more accurate and personalized recommendations~\cite{lin2023can,xi2023towards,friedman2023leveraging,bao2023tallrec}. Current efforts have to some extent integrated large language models with recommender systems, either by injecting recommendation knowledge into LLMs~\cite{bao2023tallrec,li2023ctrl,p5,zhang2021language,m6rec} or by incorporating LLM knowledge into traditional recommender models (CRMs)~\cite{xi2023towards,lyu2023llm,gong2023unified,lyu2023llm,lin2023clickprompt,wang2023flip}. However, when generating final recommendations, they adopt a binary approach, either relying entirely on LLMs or conventional recommender models. None of the previous studies have explored which specific segments of recommendation data LLMs and CRMs really excel in, neglecting potential synergies that could be leveraged to enhance recommendation quality.

\begin{figure}[t]
    \centering
% \vspace{-5pt}
    \includegraphics[width=0.8\textwidth]{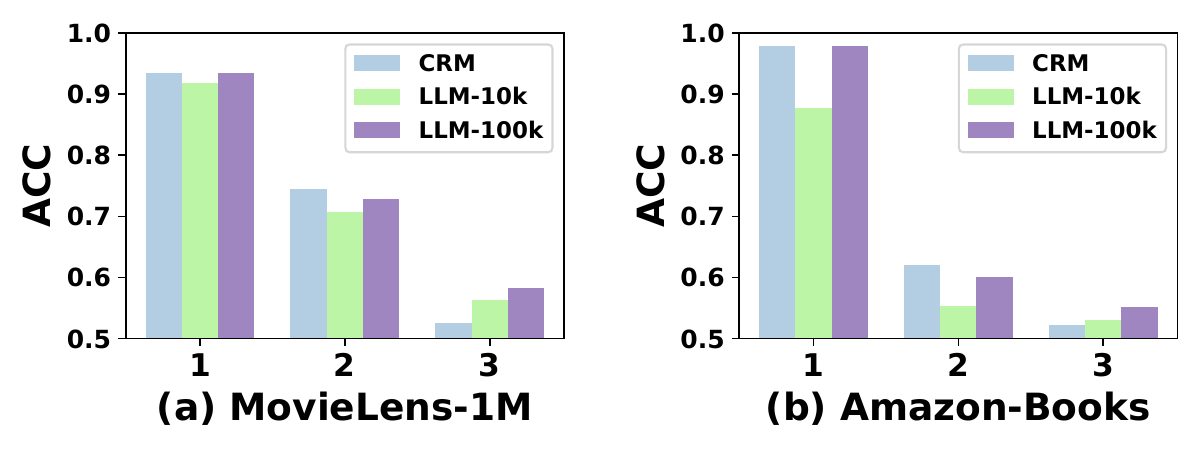}
% \vspace{-9pt}
    \caption{Performance of conventional recommender model (CRM) and Large Language Model (LLM) on different groups.}
% \vspace{-10pt}
    \label{fig:intro}
\end{figure}

In addressing this gap, we conducted experiments on two widely-used recommendation datasets, MovieLens-1M\footnote{https://grouplens.org/datasets/movielens/1m/} and Amazon-Books\footnote{https://cseweb.ucsd.edu/~jmcauley/datasets/amazon\_v2/}, comparing the performance of the representative CRM (DCNv2~\cite{DCNv2}) and LLM (LLaMA2-7B~\cite{touvron2023llama}) on different groups of data samples. First, we train a classical recommendation method DCNv2 on full training data and finetuned LLaMA2-7B with LoRA~\cite{hu2021lora} on 10k and 100k training data following similar prompt in TALLREC~\cite{bao2023tallrec}. Then, we adopt entropy as a confidence measurement~\cite{entropy}, to calculate a model's prediction uncertainty. We divided the test data into three groups by ranking DCNv2's confidence in its own predictions, with the group "\textbf{1}" having the highest confidence and the group "\textbf{3}" having the lowest, as shown in Figure~\ref{fig:intro}. Finally, we compare the performance of DCNv2 (\ie, \textbf{CRM}) and LLaMA2-7B finetuned on 10k and 100k training data, denoted \textbf{LLM-10k} and \textbf{LLM-100k} in Figure~\ref{fig:intro}, across these three groups. It is evident that in the first and second groups, LLM's results lag behind CRM, with only LLM-100k matching CRM's performance in the first group after trained on extensive data. This indicates that samples, where CRM excels, are relatively challenging for LLM, requiring substantial training data and long training time for comparable performance. This may be because a fully trained CRM can more easily capture certain collaborative signals unique to recommendations, such as associations between two items, while an LLM finetuned only on a subset of data may struggle to grasp such knowledge. On the contrary, in the third group, where CRM performs the least proficiently, both LLM-10k and LLM-100k outperform CRM. The low confidence of CRM may come from long-tail items~\cite{cleger2013being}, noisy samples~\cite{joorabloo2022improved}, polarizing items~\cite{cleger2013being}, and inconsistent user behavior~\cite{joorabloo2022improved}. On such data, LLM can leverage its extensive world knowledge, semantic understanding, and reasoning abilities to achieve better performance, even with limited training data, like 10k.

From the above findings, a straightforward idea can be easily conceived: leveraging the strengths of each model. The resource-efficient CRM, with a shorter inference time, can handle simple and moderate samples, while LLM can process a small subset of challenging samples for CRM. Here,
CRM and LLM bear a resemblance to System 1 and System 2 in the dual-process theory~\cite{groves1970habituation}, which elucidates cognitive processing mechanisms. In line with this framework, System 1 is responsible for rapid, intuitive responses to familiar and straightforward tasks, conserving cognitive resources by swiftly executing existing routines for tasks. Conversely, System 2 engages in deliberative, analytical reasoning, activated when encountering novel or demanding situations that necessitate deeper cognitive engagement. By analogy, CRM operates akin to System 1, efficiently managing straightforward recommendation tasks with established patterns. At the same time, LLM functions akin to System 2, employing its expansive knowledge and reasoning abilities to tackle complex recommendation challenges that may require deeper comprehension and analysis. Nonetheless, a notable issue arises when CRM and LLM are trained independently: their decision boundaries may diverge, that is, their boundaries between different classes or categories may be different. Merging these models without addressing this discrepancy can result in a shift in decision boundaries~\cite{wang2022augmentation}, leading to inconsistencies in how they classify or recommend items. This alteration can lead to suboptimal outcomes, undermining the efficacy of the combined approach as shown in Table~\ref{tab:ablation}.

Therefore, we propose \underline{Co}llaborative Recommendation with conventional \underline{Re}commender and \underline{L}arge \underline{La}nguage Model (dubbed \textit{CoReLLa}) wherein we engage in the joint training of the two models and alignment loss for enhanced synergy. First, LLM and CRM are trained together with a multi-stage training strategy, due to significant differences in the parameter volumes of the two models. Additionally, a specific alignment loss is devised to mitigate the issue of decision boundary shift, thereby fostering consistency in their outputs. After training, we utilize CRM's predictions to assess the difficulty level of samples and subsequently delegate challenging samples to LLM, ultimately amalgamating their outcomes. Our main contributions can be summarized as follows:
\begin{itemize}
    \item We conduct the first investigation into which subset of data CRM and LLM excel at, and find LLM performs better on data where CRM exhibits lower confidence and CRM can effortlessly handle samples that are challenging for LLM.
    \item We introduce CoReLLa, where LLM handles hard samples for CRM and addresses decision boundary shift issues through multi-stage joint training and alignment loss.
    \item Extensive experiments demonstrate that our model outperforms SOTA CRM and LLM methods significantly.
\end{itemize}

\section{Related Work}

This work is closely related to LLM-enhanced recommender systems, which can be roughly classified into two categories: (1) large language models as recommenders, and (2) conventional recommenders augmented by large language models.

\textbf{Large Language Models as Recommenders.}
As large language models (LLMs) demonstrate remarkable performance across various tasks in the field of natural language processing (NLP), researchers start to investigate the potential applications of LLMs to various recommendation tasks. One important line of methods is to adopt LLMs as recommenders to generate recommendations directly. Due to the powerful zero-shot learning and in-context learning capabilities of LLMs, early attempts primarily focus on recommendation tasks in zero-shot manners. For instance, ChatRec~\cite{chatrec} employs LLMs as recommender system interfaces for conversational multi-round recommendations. Liu \etal~\cite{liu2023chatgpt} investigate whether ChatGPT can serve as a recommender with task-specific prompts and report the zero-shot performance. Hou \etal~\cite{hou2023large} further report the zero-shot ranking performance of LLMs with historical interaction data. Sanner \etal~\cite{sanner2023large} find that LLMs provide competitive performance for pure language-based preferences in the near cold-start recommendation case in comparison to item-based CF methods.
However, directly leveraging LLMs for recommendations falls behind state-of-the-art conventional recommendation algorithms, since general-purpose LLMs lack domain knowledge and collaborative signals, which are important for recommendation tasks~\cite{lin2023can}. Therefore, the focus of later work shifts to how to inject recommendation knowledge into LLMs, primarily through parameter-efficiency finetuning. For example, TALLRec~\cite{bao2023tallrec} finetunes LLaMA-7B model~\cite{llama} with a LoRA~\cite{hu2021lora} architecture on recommendation data. ReLLa~\cite{lin2023rella} design retrieval-enhanced instruction tuning by adopting semantic user behavior retrieval as a data augmentation technique and finetunes Vicuna-13B. RecRanker~\cite{luo2023recranker} introduces instruction-tuned LLMs for diverse ranking tasks in
top-k recommendations and proposes a hybrid ranking method that ensembles various ranking tasks. 

\textbf{Conventional Recommenders Augmented by Large Language Models.}
Apart from directly adopting LLMs as recommenders, many researchers are also exploring the integration of open-world knowledge from LLMs into conventional recommendation models. Since large language models generally suffer from relatively long latency during inference, such an approach can enhance the recommendation effectiveness and meanwhile maintain the original inference efficiency, thereby avoiding the online inference latency issues caused by LLMs. For example, KAR~\cite{xi2023towards} extracts open-world knowledge from LLMs and integrates the extracted knowledge into conventional recommendation models via a hybridized expert-integrated network. LLM-Rec~\cite{lyu2023llm} designs various prompting strategies to elicit LLM's understanding of global and local item characteristics from GPT-3 (\textit{text-davinci-003}), which improve the accuracy and relevance of content
recommendations. Some researchers propose S\&R Multi-Domain Foundation model~\cite{gong2023unified}, which finetunes ChatGLM2-6B~\cite{du2022glm} to extract domain invariant features for promoting search and recommendation performance in cold-start scenarios. 

The above two types of work explore two ways of integrating LLMs and recommendations: injecting recommendation domain knowledge into LLMs and injecting LLM's knowledge into conventional recommendation models (CRM). However, regardless of which method is used, both ultimately involve using LLMs or CRM models to infer the entire dataset, without exploring whether LLMs and CRM are better suited to certain parts of the dataset. Therefore, in this work, we explore the performance of LLMs and CRMs on different parts of the dataset and allow them to leverage their respective strengths.

% \vspace{-70pt}
\section{Proposed Method}
In this work, we focus on a core task of recommender systems, Click-Through Rate (CTR) prediction, usually formulated as a binary classification problem of predicting whether a user will click on an item. The dataset is denoted as $\mathcal D = \{(x_1, y_1), \ldots, (x_i, y_i), \ldots, (x_n,  y_n)\}$, where $x_i$ represents the categorical features for the $i$-th instance, like item ID and user history, and $y_i$ denotes the corresponding binary label.

\begin{figure}[h]
    \centering
% \vspace{-5pt}
    \includegraphics[width=0.7\textwidth]{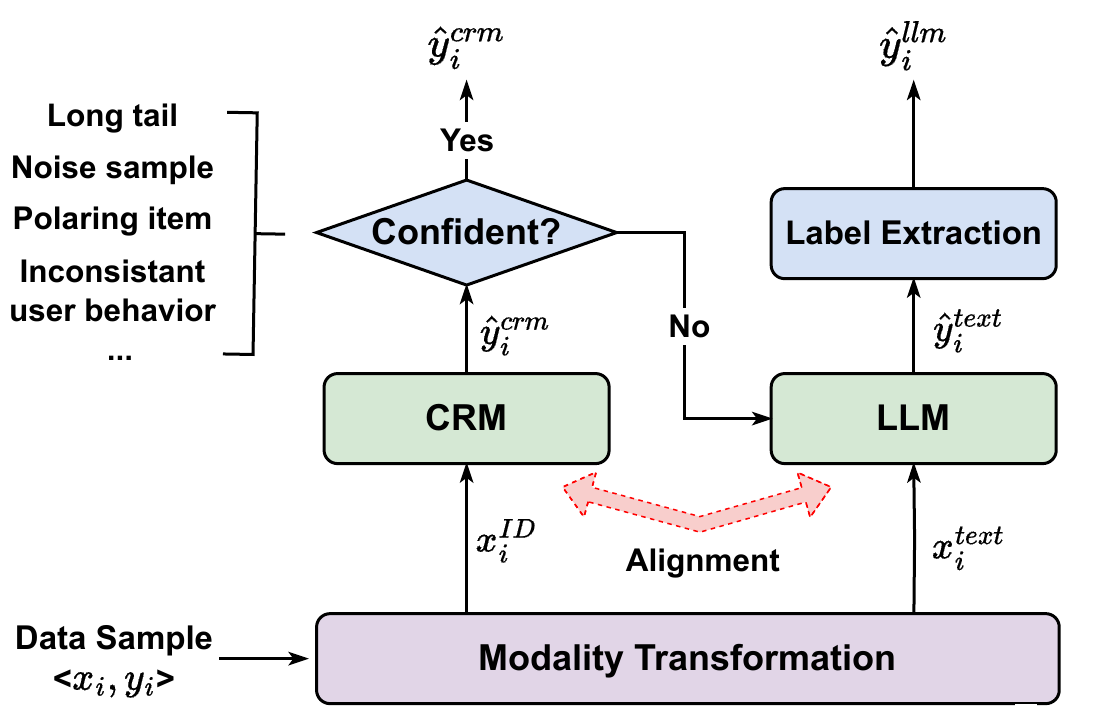}
% \vspace{-5pt}
    \caption{Framework of CoReLLa.}
% \vspace{-5pt}
    \label{fig:framework}
\end{figure}

As mentioned earlier, LLMs and CRMs each excel in different parts of recommendation data, so we have designed a framework to leverage the strengths of both CRM and LLM, making CRM handles easy samples and LLM deals with hard samples, as illustrated in Figure~\ref{fig:framework}. To mitigate decision boundary shift issues, we need to align the two models during training, for which we have designed three training stages and an alignment loss. In the inference stage,  when a sample arrives, it initially leverages the CRM branch, which has faster inference speed and lower resource consumption, to predict the result and calculate the prediction confidence. Once the confidence falls below a certain threshold, the LLM branch is activated, and the result predicted by LLM serves as the final result. Conversely, we adopt the outcome from CRM.
% It initially leverages CRM, which has faster inference speed and lower resource consumption, to predict results for all samples.\ljh{at test time, one sample per time, so we cannot say "for all samples"} Subsequently, it selects challenging samples with low confidence from CRM's predictions for further processing by LLM.\ljh{Once the prediction confidence falls below a certain threshold, we will activate the prediction branch of LLM.}\ljh{confidence threshold is a hyper, should we give?}  Finally, the predictions from both models are mixed up to obtain more accurate recommendation results. 

Specifically, a modality transformation module is introduced to transform the original data into recommendation and text modalities. For a data sample $<x_i, y_i>$, the recommendation modality input $x_i^{ID}$ for CRM is in a multi-field categorical data format, a one-hot vector. As for the text modality LLM requires, we utilize a hard template in Template~\ref{quote:prompt template}. Similarly, the binary label $y_i\in\{1,0\}$ is converted into  $y_i^{text}\in\{``\text{Yes}",``\text{No}"\}$.
\begin{equation}
  \tag{A}
  \label{quote:prompt template}
  \parbox{\dimexpr\linewidth-4em}{%
    \strut
    $\bm{x_i^{text}}=\;\,\,$``Below is the rating history of a user: \{\{user\_history\}\}. Please predict whether the user will like \{\{target\_item\}\} based on his/her rating history and the quality of the target item. You should ONLY answer no or yes. Answer: ''.
    \strut
  }
\end{equation}

Next, CRM takes $x_i^{ID}$ and generates the click probability $\hat{y}^{crm}_i$. We utilize a commonly used confidence measurement, the prediction entropy, to select hard samples, as follows: 
\begin{equation}
    s_i=-\hat{y}^{crm}_i\log \hat{y}^{crm}_i -(1-\hat{y}^{crm}_i)\log(1-\hat{y}^{crm}_i).
\label{eq:entropy}
\end{equation}
% =-\sum^{C}_{k=1}P(\hat{y}_{i,k}|x_i)\log P(\hat{y}_{i,k}|x_i)=
Typically, higher entropy indicates lower confidence in the model's predictions. The text modality $x_i^{text}$ corresponding to these hard samples is then fed into LLM. Then LLM generates the next token $\hat{y}_i^{text}$ as output, but $\hat{y}_i^{text}$ is the discrete token sampled from the distribution of LLM, not the floating-point number in $[0,1]$ required for CTR tasks. Therefore, we extract the probabilities of "Yes" and "No" from the token distribution generated by LLM, denoted as $a$ and $b$, respectively. With a bidimensional softmax, we can obtain $\hat{y}^{llm}_i$ which replaces the corresponding $\hat{y}^{crm}_i$
\begin{equation}
    \hat{y}^{llm}_i=\frac{\exp(a)}{\exp(a)+\exp(b)}\in (0,1).
\label{eq:llm_pred}
\end{equation}

Up to this point, we have only discussed the inference process of CoReLLa without delving into the training and optimization procedures. Next, we introduce a layer-wise alignment loss to facilitate the knowledge transformation between LLM and CRM, as well as aligning their outputs.
\begin{equation}
    \mathcal{L}_{cal}=\sum^n_{i=1} \sum^{C}_{j=1}\|g^{llm}(h^{llm}_{i, \mathcal{S}_j})-g^{crm}(h^{crm}_{i, \mathcal{T}_j})\|_2^{\alpha}, \,\alpha>0.
    \label{eq:alignment}
\end{equation}

In this context, both LLM and CRM consist of multiple layers, \ie, transformer blocks in LLM and cross net in CRM like DCNv2. In Equation~\ref{eq:alignment}, $h^{llm}_{i, \mathcal{S}_j}$ and $h^{crm}_{i, \mathcal{T}_j}$ denote the hidden state of the $i$-th sample at the $\mathcal{S}_j$-th and $\mathcal{T}_j$-th layers of LLM and CRM, respectively. Here, $\mathcal{S}$ and $\mathcal{T}$ are sets of layers chosen for LLM and CRM, and their size $C$ and correspondence are hyper-parameters. However, the hidden states may have different dimensions, so two transformation functions, $g^{llm}(\cdot)$ and $g^{crm}(\cdot)$, are utilized to map them into the same dimension. In practice, both $g^{llm}(\cdot)$ and $g^{crm}(\cdot)$ are a fully-connected layer. Finally, the final objective is
\begin{equation}
    \mathcal{L}=\alpha\mathcal{L}_{llm} + \beta\mathcal{L}_{crm} + \gamma\mathcal{L}_{cal},
    \label{eq:alignment}
\end{equation}
where $\mathcal{L}_{llm}$ and $\mathcal{L}_{crm}$ are the original loss of LLM and CRM, $\alpha\in[0,\infty)$, $\beta\in[0,\infty)$, and $\gamma\in[0,\infty)$ denotes the hyper-parameters that control the weight of losses.

To enhance the mix-up strategy of LLM and CRM, we employ a multi-stage training approach:
\begin{itemize}
    \item \textbf{Stage 1 (CRM warm-up training)}: In this phase, we train the CRM with the entire training set as an initialization. To achieve satisfactory results, CRM often requires substantial training data, a quantity challenging to attain during joint training with LLM. Thus, in this stage, $\alpha=\gamma=0$ and $\beta=1$. 
    \item \textbf{Stage 2 (Joint training with alignment)}: During this stage, we randomly select a small subset of training data, such as 1\%, to simultaneously train LLM and CRM while calibrating their results. Here $\alpha$, $\beta$, and $\gamma$ are non-zero. In experiments, $\alpha=\beta=1$ and $\gamma=0.1$. 
    \item \textbf{Stage 3 (LLM continue training)}: The previous stage has observed a seesaw phenomenon in the optimization of CRM and LLM --- as LLM continues to improve, CRM experiences a decline in performance. Therefore, after achieving favorable results in CRM during joint training, we cease joint training and proceed to continue training LLM with another randomly sampled subset from training data. Here, $\alpha=1$ and $\beta=\gamma=0$. 
\end{itemize}

\section{Experiments}
\subsection{Setup}
Our experiments are conducted on two public datasets, MovieLens-1M\footnote{\url{https://grouplens.org/datasets/movielens/1m/}} and Amazon-Book\footnote{\url{https://cseweb.ucsd.edu/~jmcauley/datasets/amazon_v2/}}. 
\textbf{MovieLens-1M} comprises 1 million ratings provided by 6000 users for 4000 movies. We follow common practices~\cite{DIN,qin2020user} in CTR prediction for data processing. The ratings are transformed into binary labels, with ratings of 4 and 5 labeled as positive, and the rest as negative. The data samples are sorted by their global timestamps, with the first 80\% selected as the training set, the middle 10\% as the validation set, and the final 10\% as the test set. The models receive inputs consisting of item ID, user ID, and associated attribute features of users and items.
\textbf{Amazon-Book}~\cite{ni2019justifying} is derived from the "Books" category of the Amazon Review Dataset, and it undergoes filtration to exclude less-interacted users and items. The ratings of 5 are considered positive, while the rest are deemed negative. The preprocessing procedures of Amazon-Book are akin to those applied to MovieLens-1M, with the difference being the absence of user features.

Click-Through Rate (CTR) prediction aims to predict the likelihood of a user clicking on an item, which is a core task in recommendation systems. Therefore, our experiments are conducted based on CTR prediction tasks. We select several representative traditional CTR prediction models such as DCNv2~\cite{DCNv2}, FiBiNet~\cite{FiBiNET}, AutoInt~\cite{AutoInt}, xDeepFM~\cite{xDeepFM}, Fi-GNN~\cite{Fi-GNN}, \etc, as baselines. For instance,  \textbf{xDeepFM} \cite{xDeepFM} leverages the power of both deep network and Compressed Interaction Network to generate feature interactions at the vector-wise level.   \textbf{DCNv2}~\cite{DCNv2} is an improved framework of DCN which is more practical in large-scale industrial settings. \textbf{FiBiNet}~\cite{FiBiNET} can dynamically learn the feature importance by Squeeze-Excitation network and fine-grained feature interactions by bilinear function. \textbf{FiGNN}~\cite{Fi-GNN} converts feature interactions into modeling node interactions on the graph for modeling feature interactions in an explicit way. \textbf{AutoInt}~\cite{AutoInt} adopts a self-attentive neural network with residual connections to model the feature interactions explicitly. Additionally, we also compare recommendation models based on LLM, including P5~\cite{p5}, TALLREC~\cite{bao2023tallrec}, and CTRL~\cite{li2023ctrl}, and adapted them to CTR prediction tasks. For example, \textbf{P5}~\cite{p5} is a text-to-text paradigm that unifies recommendation tasks and learns different tasks with the same language modeling objective during pretraining. \textbf{TALLRec}~\cite{bao2023tallrec} finetunes LLaMa-7B~\cite{llama} with a LoRA architecture on recommendation tasks and enhances the recommendation capabilities of LLMs in few-shot scenarios. In our experiment, we implement TALLRec with LLaMa-2-7B-chat\footnote{https://huggingface.co/meta-llama/Llama-2-7b-chat-hf}, since it has better performance and ability of instruction following. We employ widely-used \textit{ACC} (Accuracy), \textit{AUC} (Area under the ROC curve) and \textit{LogLoss} (binary cross-entropy loss) as evaluation metrics following~\cite{DCNv2,AutoInt,DIN}. A higher AUC value or a lower Logloss value, even by a small margin (\eg, 0.001), can be viewed as a significant improvement in CTR prediction performance, as indicated by previous studies~\cite{xDeepFM,DCNv2,lin2023map}.

As for our model, we opt for DCNv2~\cite{DCNv2} as the CRM and LLaMa2-7b-chat as the LLM. Firstly, DCN undergoes warm-up training on the entire dataset. Subsequently, DCN and LLaMa2-7b (finetuned with LoRA) are jointly trained on 20-30k data samples which are randomly selected from the training set. Finally, LLaMa2-7b is trained independently on the other 20-30k randomly selected data samples. Other parameters, such as batch size, learning rate, and weight decay are determined through grid search to achieve the best results. For fair comparisons, the parameters of the backbone model and the baselines are also tuned to achieve their optimal performance. 

% The detailed statistics of the processed datasets are shown in Table \ref{tab:dataset_stats}.

% \begin{table}[]
% \caption{The preprocessed dataset statistics.}
% \begin{tabular}{cccc}
% \toprule
% \textbf{Dataset} & \textbf{\#user} & \textbf{\#item} & \textbf{\#interaction} \\
% \midrule
% MovieLens-1M     & 6000            & 4000            & 1,000,209              \\
% Amazon-Books     & 11,906          & 17,332          & 1,406,582             \\
% \bottomrule
% \end{tabular}
% \label{tab:dataset_stats}
% \end{table}
\subsection{Overall Performance}
  We evaluate our proposed models and baseline models with AUC (Area under the ROC curve), ACC (Accuracy)
and LogLoss (binary cross-entropy loss) in Table~\ref{tab:overall}. Based on the experimental results, the following conclusions can be drawn: (1) our model CoReLLa significantly outperforms CRM and PLM-based models. For instance, on Amazon-Books, our proposed CoReLLa demonstrates a notable improvement over the best baselines, with a 1.38\% reduction in Logloss and a 1.03\% increase in ACC. On the MovieLens-1M dataset, CoReLLa also demonstrates an improvement of 0.72\% in AUC and 1.08\% in ACC. This indicates that CoReLLa successfully integrated the strengths of LLM and CRM models, yielding superior results than both types of models. (2) Pure PLM-based recommendations such as P5 and TALLREC often fall short compared to CRMs like FiBiNet and AutoInt. This also validates our conclusion in Figure~\ref{fig:intro} that LLM does not surpass CRM in most samples. This indicates that it remains challenging for LLM to surpass well-designed CRMs, and further integration with domain knowledge in the recommendation field may be required. However, utilizing larger language models tends to be more effective than smaller ones. For example, TALLREC based on LLaMa-7B outperforms CTRL based on BERT, suggesting that recommendation can benefit from larger language models. 
\begin{table}[]
\caption{Overall performance on two benchmark datasets. We underline the second-best value and denote the best result in bold, whose improvements are statistically significant with $p < 0.05$ against best baselines denoted by *.}
\vspace{9pt}
\centering
\scalebox{1.1}{
\setlength{\tabcolsep}{2mm}{
\begin{tabular}{ccccccc}
\toprule
\multirow{2}{*}{\textbf{Model}} & \multicolumn{3}{c}{\textbf{MovieLens-1M}} & \multicolumn{3}{c}{\textbf{Amazon-Books}} \\
\cmidrule{2-7}
 & \textbf{AUC} & \textbf{Logloss} & \textbf{ACC} & \textbf{AUC} & \textbf{Logloss} & \textbf{ACC} \\
  \midrule
DCNv2 & 0.7939 & 0.5469 & \underline{0.7230} & 0.8255 & 0.5012 & 0.7481 \\
xDeepFM & 0.7925 & 0.5449 & 0.7210 & 0.8253 & 0.5021 & 0.7481 \\
FiBiNet & \underline{0.7947} & 0.5442 & 0.7228 & 0.8254 & 0.5018 & 0.7479 \\
AutoInt & 0.7909 & 0.5472 & 0.7214 & \underline{0.8256} & \underline{0.5010} & 0.7480 \\
DeepFM & 0.7940 & \underline{0.5439} & 0.7225 & 0.8252 & 0.5015 & \underline{0.7483} \\
FiGNN & 0.7921 & 0.5464 & 0.7209 & 0.8224 & 0.5046 & 0.7458 \\
\midrule
P5 & 0.7902 & 0.5516 & 0.7174 & 0.7986 & 0.5320 & 0.7275 \\
TALLREC & 0.7931 & 0.5463 & 0.7209 & 0.8239 & 0.5060 & 0.7436 \\
CTRL & 0.7929 & 0.5465 & 0.7218 & 0.7996 & 0.5297 & 0.7253 \\
\midrule
\textbf{CoReLLa} & \textbf{0.8001*} & \textbf{0.5402*} & \textbf{0.7308*} & \textbf{0.8303*} & \textbf{0.4941*} & \textbf{0.7558*}\\
\bottomrule
\end{tabular}
}}
\label{tab:overall}
% \vspace{-4pt}
% \footnotesize \flushleft\hspace{0cm} $*$ denotes statistically significant improvement over baselines (t-test with $p$-value $<$ 0.05).
\end{table}

\subsection{Ablation Study}
\label{sec:ablation}

\begin{table}[h]
\caption{Performance of different variants on two datasets. We denote the best result in bold}
\vspace{9pt}
\centering
\scalebox{1.15}{
\setlength{\tabcolsep}{2mm}{
\begin{tabular}{ccccccc}
\toprule
\multirow{2}{*}{\textbf{Variants}} & \multicolumn{3}{c}{\textbf{MovieLens-1M}} & \multicolumn{3}{c}{\textbf{Amazom-Books}} \\
\cmidrule{2-7}
 & \textbf{AUC} & \textbf{Logloss} & \textbf{ACC} & \textbf{AUC} & \textbf{Logloss} & \textbf{ACC} \\
 \midrule
w/o S1 & 0.6511	& 0.6611 &	0.6029  & 0.8046 & 0.5303 & 0.7354 \\
w/o S2 & 0.7941 & 0.5467 & 0.7259 & 0.8265 & 0.4988 & 0.7486 \\
w/o S3 & 0.7990 & 0.5400 & 0.7284 & 0.8277 & 0.4966 & 0.7468 \\
w/o mix & 0.7982 & 0.5410 & 0.7276 & 0.8285 & 0.4959 & 0.7501 \\
CoReLLa & \textbf{0.8001} & \textbf{0.5402} & \textbf{0.7308} & \textbf{0.8303} & \textbf{0.4941} & \textbf{0.7558} \\
\bottomrule
\end{tabular}
}}
\label{tab:ablation}
% \vspace{-4pt}
\end{table}

In this section, we explore how different training stages and the mix-up strategy of CoReLLa impact the final results. We design four variants and conduct experiments on two datasets, with the results presented in Table 1~\ref{tab:ablation}. In the table, "\textbf{w/o S1}", "\textbf{w/o S2}", and "\textbf{w/o S3}" respectively denote the exclusion of training stages 1 (CRM warm-up training), stage 2 (joint training with alignment), and stage 3 (LLM continue training). "\textbf{w/o mix}" indicates generating recommendations by CRM after the joint training in stage 2 without the mix-up strategy.

From the table, we observed the most significant decrease in model performance when excluding S1, primarily due to the pivotal role of CRM in our framework. The confidence level of CRM is used to determine which samples require processing by LLM, and CRM also handles the majority of the data. Hence, the entire framework relies on a high-quality CRM model. Typically, CRMs trained on the full dataset achieve better results, and removing Stage 1 leads to poorer performance since CRMs are only trained on a small amount of data, consequently resulting in a decrease in overall model performance. The removal of S2 also results in a notable decline, even inferior to the performance of baseline CRMs, especially on AUC. This indicates that without joint training and alignment, the simple combination of LLM and CRM trained separately may experience decision boundary shifts and reduced effectiveness. While the exclusion of the mix-up strategy leads to a certain decline in performance, it still outperforms the baseline CRM. This implies that during joint training, LLM imparts knowledge to CRM, enhancing CRM's performance.

\section{Conclusion}
In this paper, we point out that current works solely use either LLM or CRM for recommendations, overlooking their distinct strengths. Therefore, we conduct the first experiments to compare the performance of CRM and LLM on various data segments. Findings show LLMs excel where CRMs exhibit lower confidence, suggesting synergies in their combination. Thus, we propose CoReLLa, which jointly trains LLM and CRM via a multi-stage training strategy and alignment loss to address the issues of decision boundary shifts. CoReLLa outperforms state-of-the-art CRM and LLM methods, highlighting its effectiveness in recommendation tasks.

\bibliographystyle{plainnat}
\bibliography{myref}

%%%%%%%%%%%%%%%%%%%%%%%%%%%%%%%%%%%%%%%%%%%%%%%%%%%%%%%%%%%%

\end{document}